\documentclass[pra,amssymb,amsmath,superscriptaddress,nofootinbib]{revtex4-2}

\usepackage{braket}
\usepackage{tikz}
\usepackage{multirow}
\usepackage{booktabs}
\usepackage[export]{adjustbox}
\usepackage{graphicx}
\usepackage{afterpage}
\usepackage{subcaption}
\usepackage[page,toc]{appendix}
\usepackage{amsthm}
\usepackage{siunitx}
\usepackage{accents}
\usepackage{dsfont}
\usepackage{hyperref}
\hypersetup{
	colorlinks=true,
	linkcolor=blue,
	filecolor=magenta,      
	urlcolor=cyan,
	citecolor=red
}
\usepackage{xcolor}
\usepackage[normalem]{ulem}
\usepackage{soul}
\usepackage{mathtools}
\DeclareMathOperator{\Tr}{Tr}
\DeclareMathOperator{\Circ}{circ}

\DeclarePairedDelimiter\abs{\lvert}{\rvert}%
\DeclarePairedDelimiter\norm{\lVert}{\rVert}%

\makeatletter
\let\oldabs\abs
\def\abs{\@ifstar{\oldabs}{\oldabs*}}
\let\oldnorm\norm
\def\norm{\@ifstar{\oldnorm}{\oldnorm*}}
\makeatother

\usepackage{afterpage}

\newcommand{\boldR}{\boldsymbol{\rho}}
\newcommand{\boldr}{\boldsymbol{r}}
\newcommand{\rt}{\boldsymbol{\rho}_t}
\newcommand{\rr}{\boldsymbol{\rho}_r}

\graphicspath{figures}

\usepackage{lineno}

\begin{document}

\title{Optimum-Transmission Free-Space Optical Communications}
\author{Prajit Dhara}
\email[]{Now at RTX BBN, Cambridge MA. Email: prajit.dhara@rtx.com}
\address{Wyant College of Optical Sciences, The University of Arizona, Tucson, AZ 85721, USA}
\address{NSF-ERC Center for Quantum Networks, The University of Arizona, Tucson, AZ 85721, USA}
\address{Department of Electrical and Computer Engineering, University of Maryland, College Park, MD 20742, USA}

\author{Babak N. Saif}
\address{NASA Goddard Space Flight Center, 8800 Greenbelt Rd, Greenbelt, MD 20771, USA}

\author{Jeffrey H. Shapiro}
\address{Research Laboratory of Electronics, Massachusetts Institute of Technology, Cambridge, MA 02139, USA}

\author{Saikat Guha}
\email[]{Email: saikat@umd.edu}
\address{Wyant College of Optical Sciences, The University of Arizona, Tucson, AZ 85721, USA}
\address{NSF-ERC Center for Quantum Networks, The University of Arizona, Tucson, AZ 85721, USA}
\address{Department of Electrical and Computer Engineering, University of Maryland, College Park, MD 20742, USA}

\email{\authormark{$\dagger$}prajit.dhara@arizona.edu \authormark{*}saikat@arizona.edu} 



\begin{abstract}

Slepian developed the Prolate Spheroidal Wavefunction (PSW) spatial-mode basis, which forms the normal modes of the Fresnel-propagation kernel of a free-space optical communications channel bookended by hard-circular apertures. The zero-th order PSW mode has the highest power-transfer eigenvalue, exciting which on the transmitter side therefore maximizes the transmissivity for single-spatial-mode communications~\cite{Slepian1964-eh,SlepianJOSA}. We show that the transmissivity performance of this fundamental PSW mode can be obtained by an aperture-truncated Gaussian beam of an optimized beam waist, despite the two mode shapes deviating from one another in the near-field regime. 
\end{abstract}

\maketitle

\section{Introduction}
\label{sec:intro}


Free-space optical (FSO) communication links are fundamentally constrained by diffraction and finite-aperture effects. Particularly, given a link propagation geometry, the maximum number of orthogonal transverse spatial modes supported by the system is constrained by the Fresnel number product ($D_f$). The foundational problem of identifying the mode set that maximizes transmissivity in such aperture-limited channels was resolved by David J.\ Slepian {\em et al.} in a series of landmark papers~\cite{Slepian1961-sl,Slepian1964-eh,Slepian1978-to,SlepianJOSA}. Specifically, the works of Slepian and co-authors implied that the normal modes of Fresnel propagation between hard circular apertures are given by two-dimensional prolate spheroidal wavefunctions (PSWs). These modes arise from maximizing the joint spatial and spatial-frequency-domain field concentration and diagonalizing the free-space propagation operator. In particular, the fundamental PSW mode uniquely maximizes the transmissivity of a single spatial mode, establishing a strict upper bound on the power that can be transferred through an aperture-limited free-space channel.

Despite their optimality, PSW modes have seen limited adoption in practical optical systems, because of their mathematical complexity, lack of closed-form expressions, and experimental challenges in their synthesis and sorting. In contrast, Gaussian beams dominate practical FSO implementations due to their simplicity, robustness, and ease of generation. This naturally raises the central question this article aims to address: how much of the link performance is sacrificed by using Gaussian beams instead of the optimal PSW modes in hard aperture-limited free-space links? 

Precise and quantitative comparison between Gaussian beams and PSW modes—particularly across near- and intermediate-field propagation regimes—has been lacking. In this work, we show that this presumed performance gap \textit{does not, in fact, exist}, {in particular in the far-field regime}. We demonstrate that an aperture-truncated Gaussian beam with an appropriately optimized waist achieves the same transmissivity as the fundamental PSW mode over a wide range of Fresnel number product. Although the spatial profiles of the two fields differ substantially in the near field, the propagation transmissivities are identical. We further show that deviations between the Gaussian and PSW field shapes manifest only as redistribution among higher-order PSW modes whose transmissivities approach unity, leaving the overall channel efficiency unchanged.

By establishing this equivalence, our results indicate that PSW modes provide a theoretical benchmark that validates the optimality of carefully chosen Gaussian beams, demonstrating that optimal single-mode transmissivity in aperture-limited free-space channels can be achieved without foregoing the ease and simplicity of standard Gaussian optics.

\section{Free Space Optical Communications with Hard Circular Apertures}
\label{sec:fresnel_diff}

\begin{figure}[h!]
	\centering
	\includegraphics[width=0.3\linewidth]{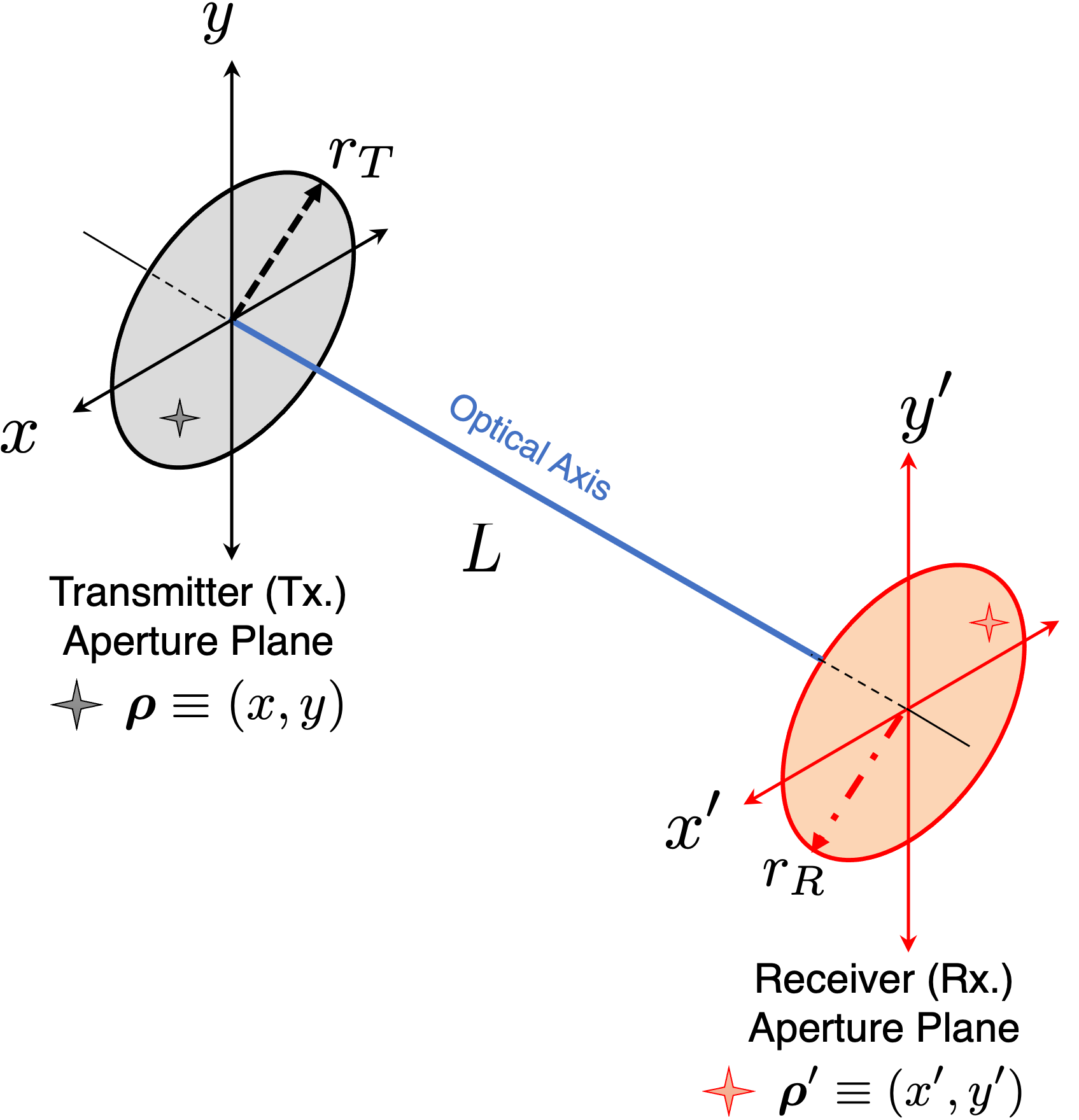}
	\caption{Hard circular aperture propagation geometry. }
	\label{fig:tradeoffs}
\end{figure} 

We begin by considering the propagation of a monochromatic optical field, $u_I(\boldR)$ of center wavelength $\lambda$ between two hard circular apertures. The field propagates from the exit pupil of a hard circular aperture $\Circ(\boldR /r_T); \boldR \equiv(x,y) $ of the transmitter situated at the transverse plane at $z=0$, to the entrance pupil of a hard-circular aperture $\Circ(\boldR' /r_R); \boldR' \equiv(x',y')$ of the receiver at the transverse plane at $z=L$, i.e., over a $L$-meter line-of-sight propagation path. Assuming the paraxial approximation holds true, the output field $u_O(\boldR')$ after truncation by the receiver aperture is given by the Fresnel propagation formula:
\begin{align}
	u_O(\boldR')=\int d^2\boldR \, u_I(\boldR) h(\boldR',\boldR),
	\label{eq:fresnel}
\end{align}
where the bi-variate impulse response of the aforesaid linear space-varying system (the Green's Function kernel) is expressed as,
\begin{align}
	h(\boldR',\boldR)= \Circ\left(\frac{\boldR'}{r_R}\right) \frac{\exp[ik(L+|\boldR-\boldR'|^2/2L)]}{i\lambda L} \Circ\left(\frac{\boldR}{r_T}\right).
	\label{eq:fresnel_kernel}
\end{align}

The transmissivity for the specified propagation configuration, given the respective input and output fields $u_I(\boldR)$ and $u_O(\boldR')$, is given by: 
\begin{align}
	\eta= \frac{\int_{|\boldR'|\leq r_R } |u_O(\boldR')|^2 d^2\boldR' }{\int_{|\boldR|\leq r_T } |u_I(\boldR)|^2 d^2\boldR}.
	\label{eq:trans}
\end{align}
A singular value decomposition of the Green's Function kernel in Eq.~\eqref{eq:fresnel} yields a modal propagation relation of the form $\Phi_{n}(\boldR) \rightarrow \sqrt{\eta_{n}} \phi_{n}\left(\boldR'\right), \; n\in\mathbb{Z}^+$. Here the input mode set  $\{\Phi_{n}(\boldR)\}$ and output mode set $\{\phi_{n}(\boldR\,^{\prime})\}$ are complete and orthonormal, over the hard circular aperture boundaries $|\boldR| \leq r_T$ and $ |\boldR'|\leq r_R $ respectively. Additionally, we choose the mode ordering (w.r.t. the index $ n $) such that the modal trasmissivities satisfy $ 1 \geq \eta_{1} \geq \eta_{2} \geq \cdots \geq 0$.

For the general treatment of Eq.~\eqref{eq:fresnel}, we may adopt a  dimensionless formulation by choosing $\boldsymbol{r} = \boldR/r_T $ and $\boldsymbol{r}' =  \boldR'/r_R$ by making the following substitutions,
\begin{subequations}
	\begin{align}
\tilde{\Phi}_{n}(\boldsymbol{r})&\equiv r_T \,\exp\left(\frac{i k|\boldr|^{2} r_T^2}{ 2 L}\right)\times  \Phi_{n}\left( r_T \boldr \right);  
\label{eq:fresnel_dimless_subs1}\\
 \tilde{\phi}_{n}\left(\boldsymbol{r}'\right) &\equiv i r_R\, \exp\left({-i k L-\frac{i k\left|\boldr^{\prime}\right|^{2} r_R^2}{ 2 L}}\right)\times  \phi_{n}\left(r_R \boldr'\right).
 \label{eq:fresnel_dimless_subs2}
	\end{align}
\end{subequations}
With this substitution, the modal propagation relation may be re-expressed as,
\begin{align}
	\sqrt{\eta_{n}} \tilde{\phi}_{n}\left(\boldr'\right)=\frac{\sqrt{D_{f}}}{\pi} \int_{|\boldr| \leq 1} d^{2} \boldr \, e^{-i 2 \sqrt{D_{f}} \boldr \cdot \boldr'}\, \tilde{\Phi}_{n}(\boldr), 
	\label{eq:fresnel_dimless}
\end{align}
where $ D_{f}=\left(\pi r_T r_R /  \lambda L\right)^{2}$ is the free-space Fresnel number product for the propagation configuration. 

For the normal modes that exhibit circular symmetry, we may use $\tilde{\Phi}_{1}(\boldr)=\tilde{\Phi}_{1}(\rho)$, where $r=|\boldr|$, thereby simplifying the modal input-output relation to
\begin{align}
	\sqrt{\eta_{n}} \tilde{\phi}_{1}\left(r^{\prime}\right)=2 \sqrt{D_{f}} \int_{0}^{1}  dr\; r \, \tilde{\Phi}_{n}(r) \times J_{0}\left(2 \sqrt{D_{f}} r^{\prime} r \right). 
	\label{eq:fresnel_dimless_rad}
\end{align}

\section{Prolate Spheroidal Wavefunctions: Review}
\subsection{Prolate Spheroidal Wavefunctions on the Real Line and Extensions}
\label{subsec:PSW_1d}
{
We consider the class of all square-integrable complex-valued functions $f(t)$ of a single variable $t$ (henceforth referred to as `time') defined on the real line, $t\in(-\infty,\infty)$; we represent this class as $\mathcal{L}^2_{\infty}$. The \textit{total energy} of the function is given by $\norm*{f(t)}^2_\infty = \int_{-\infty}^{\infty} |f(t)|^2 dt$. The \textit{time-limited energy} of the function $f(t)$ in the interval $t\in(-A,A)$ is denoted by $\norm*{f(t)}^2_{A} $, and defined as,
\begin{align}
\norm*{f(t)}^2_{A}=\int_{-A}^{A} |f(t)|^2 dt.
\end{align}
Correspondingly, the class of square integrable complex-valued functions $f(t)$ defined over the interval $t \in (-A,A)$ is denoted by $\mathcal{L}^2_{A}$. 

We shall also use the Fourier transform of $f(t)$, which we represent by $F(\omega)$, where $\omega$ is the `angular frequency' variable. Any function $f(t)\in\mathcal{L}^2_{\infty}$ and its Fourier transform are related to each other by the standard Fourier relations,
\begin{align}
	\begin{split}
	    f(t)&=\frac{1}{2\pi}\int_{-\infty}^{\infty} F(\omega) e^{i\omega t} d\omega, \\
	F(\omega)&=\int_{-\infty}^{\infty} f(t) e^{-i\omega t} dt .
	\end{split}
\end{align}

For the development of the prolate spheroidal wavefunction basis, we consist some special subspaces of $\mathcal{L}^2_\infty$. First, we consider the class of \textit{band-limited} functions, $\mathfrak{B}$, whose Fourier transforms vanish for $|\omega|>\Omega$, i.e., given a function $f(t)\in\mathfrak{B}$, the Fourier transform $F(\omega)=0\; \forall \,|\omega|>\Omega$. Hence by definition, all members of $\mathfrak{B}$ satisfy the band-limited inverse Fourier relation,
\begin{align}
		f(t)&=\frac{1}{2\pi}\int_{-\Omega}^{\Omega} F(\omega) e^{i\omega t} d\omega.
\end{align}
Given an arbitrary function $g(t)\notin \mathfrak{B}$ with a Fourier transform $G(\omega)$, we can always construct its band-limited analogue, in terms of the band-limited inverse Fourier transform of $G(\omega)$, denoted by $B_\Omega[g(t)]$), and defined as,
\begin{align}
	B_\Omega[g(t)]=\frac{1}{2\pi}\int_{-\Omega}^{\Omega} G(\omega) e^{i\omega t} d\omega.
\end{align} 
Additionally, we consider the class of \textit{time-limited} functions $\mathfrak{D}$ of $\mathcal{L}^2_\infty$ which satisfy $f(t)=0\,\forall \, |t|>T$. Similar to the construction of the band limited analogiues, we can always construct a time-limited analogue of an arbitrary function $h(t)\notin \mathfrak{D}$, denoted by $D_T[h(t)]$ and defined as,
\begin{align}
	D_T[h(t)] =\begin{cases}
		h(t) &; |t|\leq T\\
		0 &; |t|>T
	\end{cases}
\end{align}

It is a well-known fact in harmonic analysis that a function cannot simultaneously be time and band-limited, i.e., an arbitrary function $f(t)\in\mathcal{L}^2_\infty$ can either be a member of $\mathfrak{B}$, or $\mathfrak{D}$ but never both. A natural question that might then arise is whether one can find an orthogonal set of band-limited functions that are simultaneously orthogonal for a specified temporal band. In a series of papers~\cite{Slepian1961-sl,Landau1961-vr,Landau1962-lj,Slepian1964-eh}, Slepian, Landau, and Pollack studied this very question. The main findings of these works established the prolate spheroidal wavefunction basis. More specifically, Refs.~\cite{Slepian1961-sl,Landau1961-vr,Landau1962-lj,Slepian1964-eh} highlight some key results for \textit{effectively time and bandwidth limited} functions of a single variable that we summarize below.

Given a finite positive temporal bandwidth of $T$ and spectral bandwidth of $\Omega $, one can find a countably infinite set of real functions $\psi_k(t); k\in\{0,1,2,\ldots\}$ with a corresponding \textit{ordered} set of real numbers $\lambda_k$ (i.e.\ $\lambda_{k+1}\leq \lambda_k \; \forall\, k\geq0$). The set of functions $\{\psi_k(t)\}$ is commonly called the one-dimensional prolate spheroidal wave-functions (PSW), and they satisfy the following properties:
\begin{enumerate}
	\item The PSWs $\{\psi_k(t)\}$ form a complete orthonormal set of functions for $t\in(-\infty,\infty)$, i.e.\,
	\begin{align}
		\int_{-\infty}^{ \infty} \psi_i(t) \psi_j(t) dt = \delta_{i,j}
	\end{align}
    By definition, their Fourier transforms $\{{\Psi}_k(\omega)\}$ form a complete orthonormal set of functions over $\omega\in[-\Omega,\Omega]$.
    
	\item Elements of $\{\psi_k(t)\}$ are complete and orthogonal for time-limited interval $t\in[-T/2,T/2]$, 
	\begin{align}
		\int_{-T/2}^{T/2} \psi_i(t) \psi_j(t) dt = \lambda_i \delta_{i,j}
	\end{align}
    
	\item For all values of $t$, the time-limited version of $\psi_i(t)$ satisfies the eigenvalue relation
	\begin{align}
		\lambda_i\, \psi_i(t) = \int_{-T/2}^{T/2} \frac{\sin \Omega(t-s)}{\pi (t-s)} \psi_i(s) ds
        \label{eq:time_limited_energy}
	\end{align}
\end{enumerate}

In particular, the integral in  Eq.~\eqref{eq:time_limited_energy} can be expressed in the more recognizable form,
\begin{align}
    \int_{-T/2}^{T/2} \frac{\sin \Omega(t-s)}{\pi (t-s)} \psi_i(s) ds &= \int_{-\infty}^{\infty} \frac{\sin \Omega(t-s)}{\pi (t-s)} \psi_i(s) \mathrm{rect}\left(\frac{s}{T/2}\right) ds\\
    &=\frac{\Omega}{\pi} \mathrm{sinc}(\Omega t)\,\ast\, \left[ \psi_i(t) \, \mathrm{rect}\left(\frac{t}{T/2}\right)\right]\\
    &= \frac{\Omega}{\pi} \mathcal{F}^{-1} \left[\mathrm{rect}\left[\frac{\omega } {\Omega}\right]\mathcal{F}\left[ \psi_i(t) \, \mathrm{rect}\left(\frac{t}{T/2}\right)\right]\right]
\end{align}
where $\mathcal{F},\mathcal{F}^{-1}$ are shorthands for Fourier transform and its inverse.
Hence, one may conclude that the eigenvalue relations account for both temporal and spectral bandwidths (applied by the corresponding rect functions). This asserts that for all $f(t)\in\mathcal{L}^2_\infty$, the class of functions that maximize the ratio of the time-limited and band-limited energy, $\norm*{D_{T/2}\circ B_{\Omega}[f(t)]}^2_\infty\equiv\norm*{B_\Omega[f(t)]}^2_{T/2}$ to the total energy $E=\norm*{f(t)}^2_\infty$ are the PSW functions.
}

\subsection{Optical Communication with Hard Apertures: 2D Generalization of the PSW Basis}
\label{subsec:PSW_2d}

The two-dimensional generalization of PSWs is quite relevant to the study of free-space optical propagation~\cite{SlepianJOSA, Slepian1964-eh}. We again consider the class of arbitrary complex-valued functions that are square-integrable over the real plane (parameterized by the real vector $\boldR\equiv(x,y)$). The conjugate space of $\boldR$ is parameterized by the vector $\boldsymbol{\sigma}$. The findings of Slepian in Ref.~\cite{Slepian1964-eh,SlepianJOSA} states that we can determine a set of band-limited functions $\{\tilde{\psi}_{k,l}(\boldR)\}$ with $ k,l \in \{0,1,2,3,\ldots\}$) and a set of corresponding positive real numbers $\tilde{\lambda}_{k,l}$ which exhibit the following properties --
\begin{enumerate}
	\item $\{\tilde{\psi}_{k,l}(\boldR)\}$ forms a complete orthonormal set of functions for all $|\rho|<\infty$ satisfy the relation
	\begin{align}
		\int_{\mathrm{all }\boldR} d^2\boldR \; \tilde{\psi}_{k,l}(\boldR) \tilde{\psi}_{k',l'}(\boldR) =\delta_{k,k'} \delta_{l,l'}
	\end{align}
    By definition, their Fourier transforms $\{\tilde{\Psi}(\sigma)\}$ form a complete orthonormal set of functions over $|\sigma|<c'$.
    
	\item The $\tilde{\psi}_{k,l}(\boldR)$ are complete and orthogonal over a space-limited region $|\boldR|\leq c; c\in\mathbb{R}^+$
	\begin{align}
		\int_{|\boldR|\leq c} d^2\boldR \; \tilde{\psi}_{k,l}(\boldR) \tilde{\psi}_{k',l'}(\boldR) =\tilde{\lambda}_{k,l}\, \delta_{k,k'} \delta_{l,l'}
	\end{align}
	\item The space-limited version of $\tilde{\psi}_{k,l}(\boldR)$, satisfies the two-dimensional eigenvalue relation
	\begin{align}
		\tilde{\lambda}_{k,l}  \tilde{\psi}_{k,l}(\boldR) = \int_{|\tilde{\boldR}|\leq c}  d^2\tilde{\boldR} \; e^{i\boldR\cdot\tilde{\boldR}}\;  \tilde{\psi}_{k,l}(\tilde{\boldR})
	\end{align}
\end{enumerate}
Analogous to time and band-limited energy maximization for one-dimensional functions, it is straightforward to show that the 2D PSW functions satisfy an equivalent role over the real plane -- they are the functions that maximize energy concentration in space-limited and spatial frequency-limited regions simultaneously. 

The Fresnel diffraction kernel for free space propagation is similar to the two-dimensional Fourier relation  between the input and output plane (up to a complex phase factor). For the specific case of Fresnel diffraction with hard apertures (i.e. kernel presented in Eq.~\ref{eq:fresnel_kernel}), the action of hard apertures on the transmitter side ($|\boldR|<r_T$) and the receiver side ($|\boldR'|<r_R$) is equivalent to space and band-limiting over the real plane. Correspondingly, the two-dimensional generalization of the PSW functions provides the necessary toolset for analyzing a complete mode set for the hard-aperture propagation kernel. Formally the 2D-PSW basis is introduced by studying the solutions for eigenvalue relation~\cite{Slepian1964-eh,SlepianJOSA},
\begin{align}
	\int_{y_1^2+y_2^2\leq1 } \exp({i g \left(x_{1} y_{1}+x_{2} y_{2}\right)})\; \psi\left(y_{1}, y_{2}\right) d y_{1} d y_{2}= \alpha \,\psi\left(x_{1}, x_{2}\right).
	\label{eq:slepian_rect}
\end{align} 
This is similar to the dimensionless formulation of the standard Fresnel diffraction equation( Eq.~\eqref{eq:fresnel}), upto a complex phase factor of $\exp(i k(|\boldR|^2+|\boldR'|^2)/2L)$, where $ g\equiv2\sqrt{D_f} =2\pi r_T r_R /\lambda L$ and $\alpha\equiv \sqrt{\eta}/(\sqrt{D_f}/\pi) $. Transformation of the eigenvalue relation to a radial coordinate system, $ (x_1,x_2)\rightarrow (r,\theta); (y_1,y_2)\rightarrow(r',\theta') $, yields an alternative definition of the 2D-PSW basis, 
\begin{align}
    \alpha\, \psi(r, \theta) &=\int_{0}^{1} d r^{\prime} r^{\prime} \int_{0}^{2 \pi} d \theta^{\prime} e^{i {grr}^{\prime} \cos \left(\theta-\theta^{\prime}\right)} \psi\left(r^{\prime}, \theta^{\prime}\right) \nonumber\\
    &=\sum_{-\infty}^{\infty} i^{m} e^{i m \theta} \int_{0}^{1} d r^{\prime} r^{\prime} J_{m}\left({grr}^{\prime}\right) \int_{0}^{2 \pi} d \theta^{\prime} e^{-i m \theta^{\prime}} \psi\left(r^{\prime}, \theta^{\prime}\right).
    \label{eq:radial_PSW}
\end{align}
where Ref.~\cite{Slepian1964-eh} finds that Eq.~\eqref{eq:radial_PSW} admits solutions of the form,
\begin{subequations}
	\begin{align}
		\psi_{0, n}(r, \theta)&=R_{0, n}(r),\quad  n=0,1,2,\dots,  \\
		\psi_{\ell,p}(r, \theta)&=\begin{cases}
			R_{\ell,p}(r) \cos \ell\theta
			\\R_{\ell,p}(r) \sin \ell\theta  
		\end{cases}\;N=0,1,2,\ldots; n=0,1,2,\ldots,
	\end{align}
\end{subequations}
where $\ell,p$ are the mode indices and $ R_{\ell,p}(r)$ must satisfy the equation,
\begin{align}
	\beta_{\ell,p} R_{\ell,p}(r)=\int_{0}^{1} J_{\ell}\left(g r r^{\prime}\right) R_{\ell,p}\left(r^{\prime}\right) r^{\prime} d r^{\prime},
	\label{eq:PSW_rad}
\end{align}
with $ \alpha_{\ell,p}=2 i^N \pi \beta_{\ell,p} $. { Readers may note that we replace Slepian's mode index notation of $(N,n)$ with $(\ell,p)$.} By making the substitutions $\gamma_{\ell,p}=\sqrt{g} \beta_{\ell,p}$ and $\varphi_{\ell,p}(r)=\sqrt{r} R_{\ell,p}(r)$
the symmetric equation for the PSW basis is defined by,
\begin{align}
	\gamma_{\ell,p} \, \varphi_{\ell,p}(r)=\int_{0}^{1} J_{\ell}\left(g r r'\right) \sqrt{g r r'}\, \varphi_{\ell,p}\left(r^{\prime}\right) d r^{\prime}.
	\label{eq:PSW_symm}
\end{align}
Deriving an explicit form for the 2D PSW basis is non-trivial -- the functions are typically expressed as convergent sums of Bessel functions~\cite{Slepian1964-eh}, with the general form, 
\begin{align}
		\varphi_{\ell,p}(r)=\frac{1}{\gamma_{\ell,p}} \sum_{j=0}^{\infty} d_{j}^{\ell,p} \frac{J_{\ell+2 j+1}(g r)}{ \binom{\ell+j}{j} \sqrt{g r}},
	\label{eq:PSW_expansionBessel}
\end{align}
where $J_{n} (\cdot) $ is the Bessel function of the $ n $-th order and $ d_{j}^{\ell,p} $ are coefficients of the expansion, which form a three-term recurrence relation when substituted in the eigenvalue relation. We provide detailed calculations of the coefficients, using a method for solving recurrence relations introduced in~\cite{Bouwkamp1947-no}) in the Supplemental Material~\ref{app:slepian_defs}. 

\begin{figure}[ht!]
	\centering
	\includegraphics[width=0.6\linewidth]{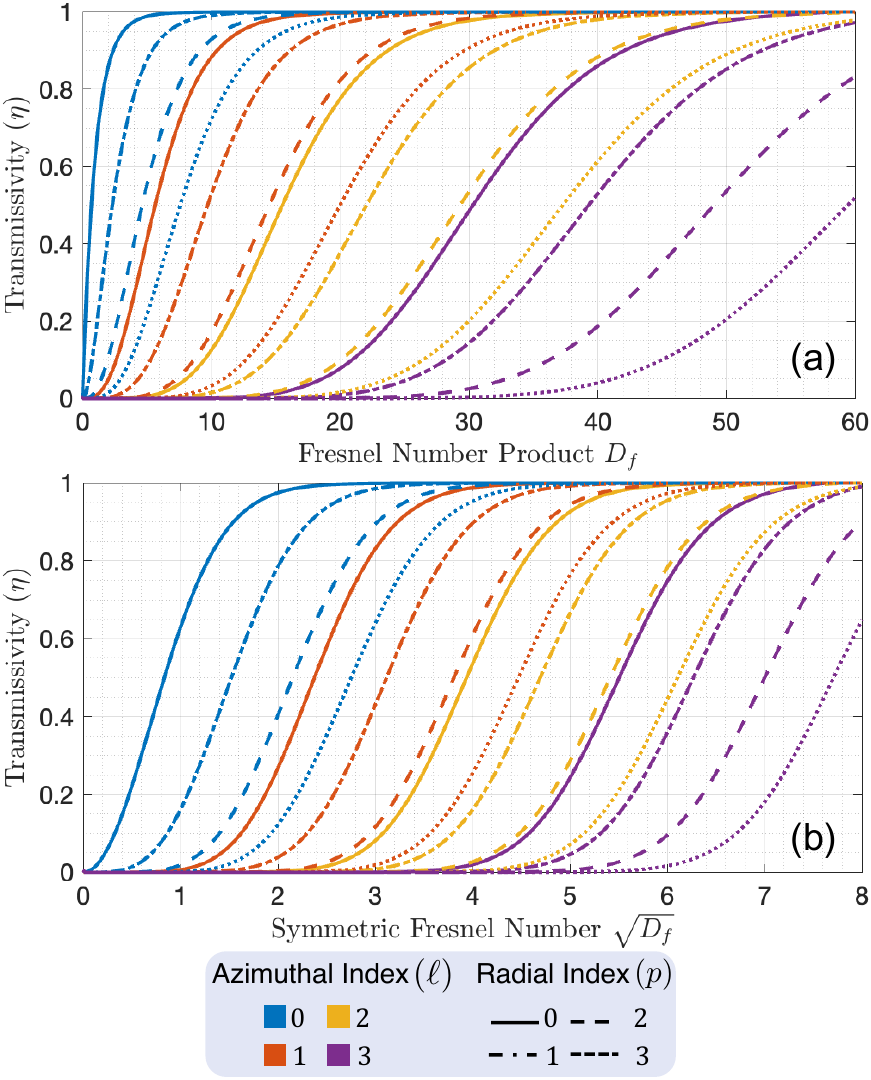}
	\caption{Plot of modal transmissivity $ \eta $ of the PSW modes as a function of the free-space Fresnel number product $ {D_f} $ (a) and the symmetric Fresnel number ($\sqrt{D_f}$). The mode indices are indicated in the legend.}
	\label{fig:eta_vs_Df}
\end{figure}

The modal transmissivities of the PSW mode set w.r.t. $ D_f$ are shown in Fig.~\ref{fig:eta_vs_Df} (a). It is common to represent the transmissivity as a function of $ \sqrt{D_f} =g/2$ as is shown in Fig.~\ref{fig:eta_vs_Df} (b) -- { in this representation it is easy to see that for a specified value of $D_f$ there are  $\sim{D_f}$ modes with transmissivity $\eta\rightarrow 1$.}
	
\begin{figure}[ht!]
	\centering 
	\includegraphics[width=0.6\linewidth]{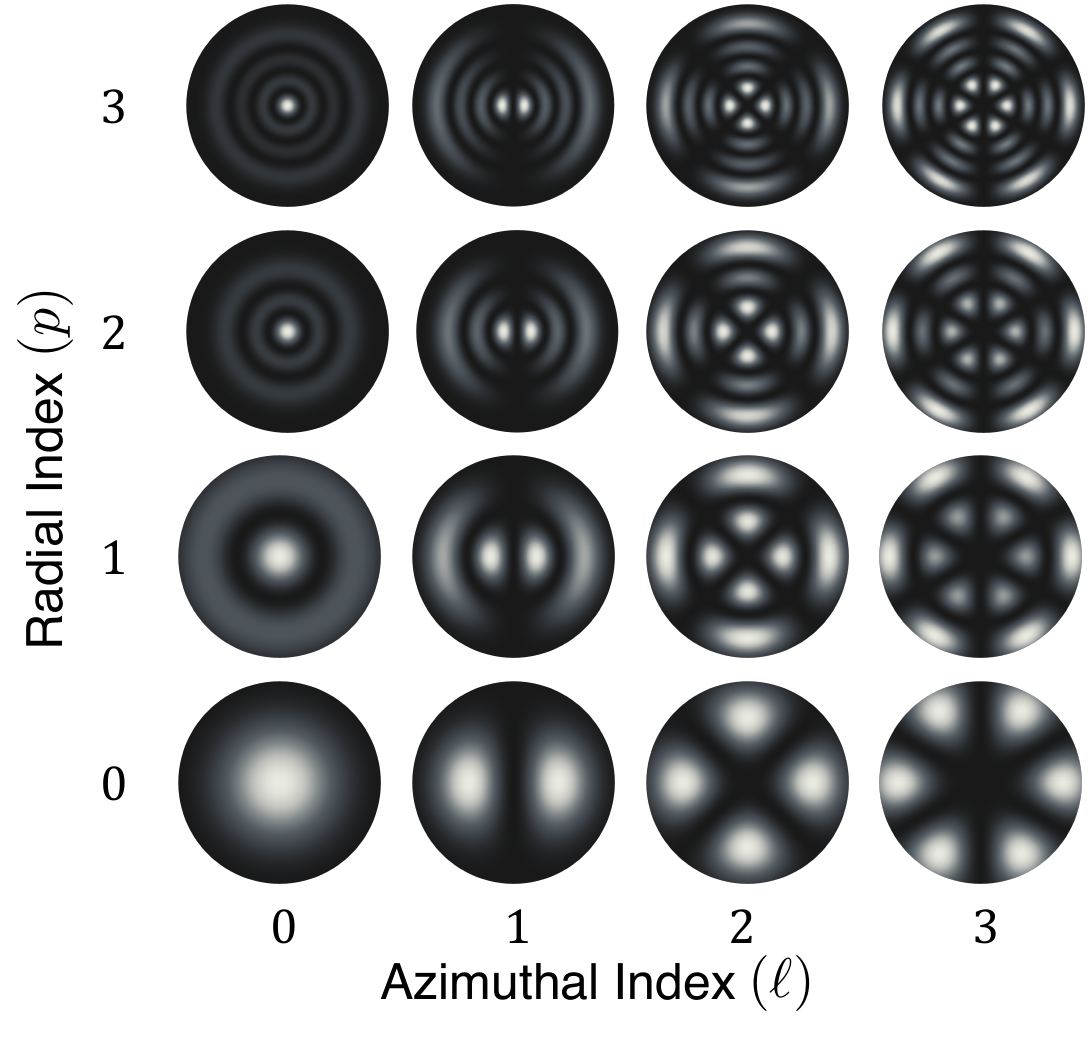}
	\caption{Plot of higher order PSW modes $\psi_{\ell,p}(r,\theta)$ for $\ell=\{0,1,2,3\}$ and $p=\{0,1,2,3\}$. For each mode subplot, we choose an appropriate $D_f$ parameter (see Table~\ref{tab:psw_mode_df}) to capture the relevant features of the corresponding mode. (i.e.\ the angular and radial nodes).}
	\label{fig:slep_modes}
\end{figure}

{ 
	We plot the PSW modes for $N=\{0,1,2,3\}$ and $n=\{0,1,2,3\}$ in Fig.~\ref{fig:slep_modes} -- readers should note that the $D_f$ parameter has been chosen for each mode so as to capture the relevant features of the corresponding mode (tabulated in  Table~\ref{tab:psw_mode_df}). We note that the $(\ell,p) $-th mode has $n$ radial nodes and $2N$ angular nodes. The radial component of the generalized PSW mode $\psi_{\ell,p}(r,\theta)$,  represented by $\varphi_{\ell,p}(r)$ (ref. Eq.~\eqref{eq:PSW_expansionBessel}) has $n$ zero crossings in the normalized aperture region $r\leq 1$. The angular modulation function of $\psi_{\ell,p}(r,\theta)$ i.e.\ $\sin(\ell\theta)$ or $\cos(\ell\theta)$ has $2N$ zeros for $\theta\in[0,2\pi]$.

    Readers should note that although the field distributions resemble the field distribution waveforms of the Laguerre-Gauss (LG) mode set, the PSW mode set is mathematically distinct. Primarily, the LG waveforms (say $u_{\ell,p}(\boldR)$) are eigenfunctions of the Gaussian apodized Fresnel diffraction kernel of the form,
    \begin{align}
	h_{\rm LG}(\boldR',\boldR)= \exp\left(\frac{-\boldR'^2}{2r_R^2}\right) \frac{\exp[ik(L+|\boldR-\boldR'|^2/2L)]}{i\lambda L} \exp\left(\frac{-\boldR^2}{2r_T^2}\right),
	\label{eq:fresnel_gauss_kernel}
    \end{align}
    The eigenvalue relation (similar to Eq.~\eqref{eq:fresnel_dimless_gauss}) for the Gaussian apodization kernel is given in the dimensionless coordinates as, 
    \begin{align}
	\sqrt{\eta_{\ell,p}} \tilde{U}_{\ell,p}\left(\boldr'\right)=\frac{\sqrt{D_{f}}}{\pi} \int_{\boldr\in\mathbb{R}^2} d^{2} \boldr \,e^{-|\boldr|^2}\, e^{-i 2 \sqrt{D_{f}} \boldr \cdot \boldr'}\, \tilde{u}_{\ell,p}(\boldr), 
	\label{eq:fresnel_dimless_gauss}
    \end{align}
    As a result, the LG mode set has an implicit requirement that $\lim_{|\boldr|\rightarrow\infty}{\tilde{u}}_{\ell,p}(\boldr)=0$, to ensure proper orthonormality over the \textit{entire exit pupil plane of the transmitter}. In contrast, there are no convergence requirements for the PSW modes as $|\boldr|\rightarrow\infty$, since the hard circular apertures naturally truncate the wavefunctions. 

    Our previous observation indicates that both the fundamental modes for the PSW mode set $(\psi_{0,0}(\boldr))$ and the LG mode set $(u_{0,0}(\boldr)$ have no radial or angular nodes. We explore the equivalence of the fundamental mode functions in the following chapter.
}

\begin{table}[t]
  \centering

  \setlength{\tabcolsep}{6pt}
  \renewcommand{\arraystretch}{1.4}

  \begin{tabular}{
      c | 
      S[table-format=1.0]
      S[table-format=3.0]
      S[table-format=3.0]
      S[table-format=3.0]
      S[table-format=3.0]
  }
    \toprule
    \multicolumn{1}{c}{}
    & \multicolumn{1}{c}{}
    & \multicolumn{4}{c}{Angular mode index ($\ell$)} \\
    \cmidrule(lr){3-6}
    \multicolumn{1}{c}{}
    & \multicolumn{1}{c}{}
    & {0} & {1} & {2} & {3} \\
    \midrule
    \multirow{4}{*}{\rotatebox[origin=c]{90}{\shortstack[c]{Radial mode\\ index ($p$)}}}
    & 0 & 5 & 10 & 10 & 10 \\
      & 1 &  10 &  25 &  25 &  25 \\
      & 2 &  60 & 100 &  50 &  50 \\
      & 3 & 100 & 120 & 120 & 100 \\
    \bottomrule
  \end{tabular}
  \caption{Fresnel number product ($D_f$) used in the modal shape plots shown in Fig.~\ref{fig:slep_modes}}
  \label{tab:psw_mode_df}
\end{table}

\section{Optimal fundamental mode transmissivity achieving Gaussian spatial function }
\label{sec:opti_gauss}

In order to examine any equivalence between the fundamental PSW and LG mode functions, we ask if the modes yield equivalent eigenvalues under the eigenrelation of Eq.~\eqref{eq:fresnel_dimless}. This is equivalent to comparing the transmissivity of the optimal LG fundamental mode with that of the PSW fundamental mode. We consider the most general normalized form of the fundamental LG wavefunction,
\begin{align}
		\tilde{\Psi}_{\mathrm{in}} (\xi; \xi_0)= \frac{e^{-\xi^2/4\xi_0^2}}{\sqrt{2\pi \xi_0^2 (1-e^{-1/2\xi_0^2})}}.
	\label{eq:trunc_gauss}
\end{align}
where by definition, $\int_0^{1} d\xi \, \xi |\tilde{\Psi}_{\mathrm{in}}(\xi;\xi_0)|^2 =1$. Since (by definition) the waveform has no Laguerre polynomial factor in the radial direction, we will simply refer to this as a Gaussian waveform. For a given value of $ D_f $ (determined by the propagation geometry), we numerically evaluate the optimal value of $ \xi_0 $ that maximizes the transmissivity of the propagating beam, when hard circular apertures are considered at both transmitter and receiver pupil planes.

\begin{figure}[h!]
	\centering
	\includegraphics[width=0.6\linewidth]{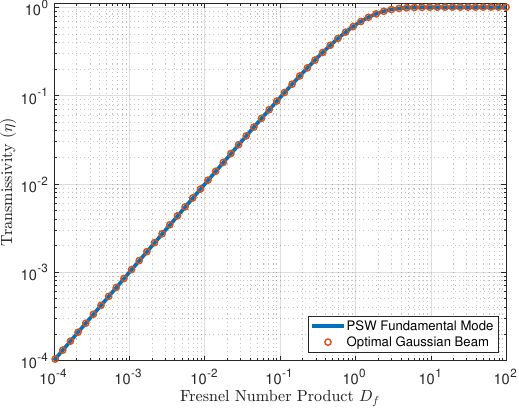}
	\caption{Comparison of optimal (numerically optimized $ \xi_0 $ in Eq.~\eqref{eq:trunc_gauss}) Gaussian input field's transmissivity (red circles) with $ \eta$ of the fundamental PSW mode $ \varphi_{0,0} $ (blue) w.r.t. the Fresnel number product $ D_f $. }
	\label{fig:compareGauss}
\end{figure}

\begin{figure}[h!]
	\centering
	\includegraphics[width=0.6\linewidth]{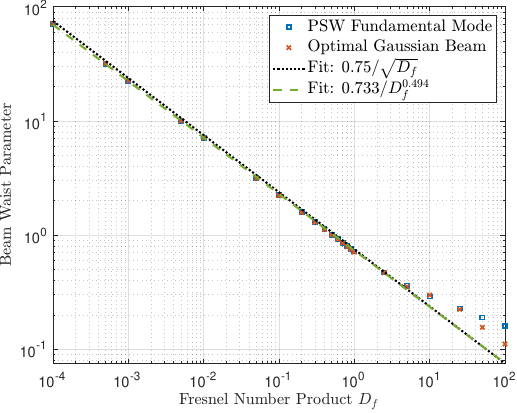}
	\caption{Numerically obtained optimal beam parameter ($ \xi_M $) plotted as a function of $ D_f $ (blue squares). We note the approximate parameter fit: $ \xi_M\approx {0.75}/{\sqrt{D_f}}$ (black dotted), and numerical fit $ \xi_M \approx 0.733/(D_f)^{0.494}$ (green dashed).  The extracted $ \xi $ parameter from the corresponding $ \varphi_{0,0} $ mode (red cross marks) is also shown. }
	\label{fig:optibeam_fit}
\end{figure}

We plot the transmissivity of the optimal Gaussian beam vs. $\sqrt{D_f}$ (red circle markers) along with the transmissivity of the fundamental PSW mode (blue line) in Fig.~\ref{fig:compareGauss}. We observe that the transmissivity curves overlap closely, indicating that the optimal Gaussian beam is indeed equivalent to the fundamental PSW mode. Additionally, we plot the $\xi_M$ parameter vs.\ $D_f$ in Fig.~\ref{fig:optibeam_fit} .  We note the approximate relation $	\xi_M (D_f)\approx{0.75}/{\sqrt{D_f}}$  (red dashed line). { A more accurate power law fit $	\xi_M (D_f)\approx{0.733}/{{D_f}^{0.494}}$ (blue solid line) is obtained by standard curve fitting routines.} We note the departure of the optimal beam parameter from the fits in the near-field, i.e., $ D_f\gtrsim5 $; this is also the region where the optimal beam waist parameter is less than 0.5. For an equal-footed comparison of the beam waists, a `waist' parameter for the $ \varphi_{0,0} $ mode is extracted (light blue circles) and plotted to highlight the key differences. Here the `waist' parameter corresponds to the value of $ \xi_0$ (say $ \xi'_M $) that maximizes the overlap integral $	\int_0^{1} d\xi \, \xi \tilde{\Psi}_{\mathrm{in}}(\xi;\xi_0)\times \varphi_{\ell,p}(\xi)$ for a given value of $D_f$.

\begin{figure}[h!]
	\centering
	\includegraphics[width=0.75\linewidth]{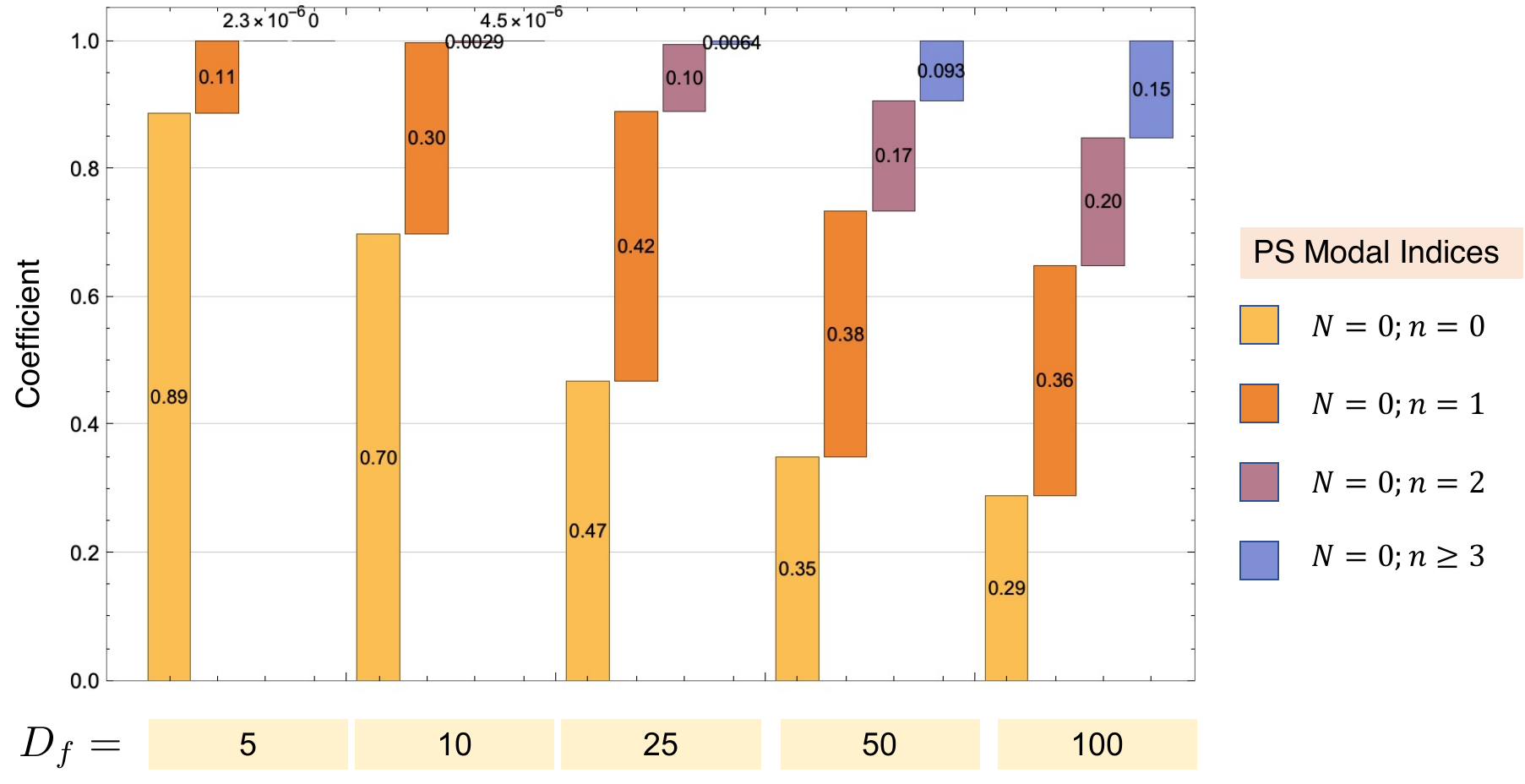}
	\caption{Mode occupation statistics of the optimal Gaussian beam in the various PSW radially symmetric modes  (i.e. $\varphi_{0,n} (x);  n=0,1,2,\ldots $) in the near-field. The modal indices are represented by the colored bars (indices in figure legend), and the value of $ D_f $ is marked on the horizontal axis. The value of the coefficients is shown on the individual bars; the sum of the coefficients must be unity.}
	\label{fig:mode_occupation}
\end{figure}

It is interesting to note that $ \xi'_M <\xi_M$ for $ D_f \geq5 $; meaning the optimal Gaussian beam does not match the PSW fundamental mode in this regime and \textit{underfills} the aperture. One may obtain the overlap of the Gaussian beam with higher-order PSW modes to understand the mode occupation. Given the radial symmetry and absence of angular nodes (i.e.\ where the function becomes zero) of the candidate Gaussian beam, we consider the circular symmetric PSW modes, i.e.\ $ \varphi_{\ell,p}(x) $ with $N=0$. We evaluate the overlap integral of the optimal Gaussian beam to these higher modes,
\begin{align}
	\langle \tilde{\Psi}_{\mathrm{in}}(r;\xi_0),\psi_{0,n}(r,\theta) \rangle =2\pi \int  r\, dr\, \tilde{\Psi}_\mathrm{in}(r;\xi_0) \times \psi_{0,n}(r,\theta).
\end{align}
Fig.~\ref{fig:mode_occupation} depicts the mode overlap coefficients of the optimal Gaussian beam for a propagation geometry of $D_f=\{5,10,25,50,100\}$ (marked on the horizontal axis). The different-colored bars represent the overlap integral values (mode indices in the legend) with the PSW mode, with $N=0$ coefficients, for $n=0,1,2$, and fill the residual in the $ n\geq3$ bin. The reader may note that as the value of $ D_f $ increases, the occupation of the higher-order modes increases. Since the transmissivities of higher order modes become essentially 1, i.e., $\eta_{0,k}\rightarrow 1; k\geq1$ (refer to Fig.~\ref{fig:eta_vs_x}), the net propagation transmissivity is unaffected; despite the deviation of the field-amplitude profiles of the fundamental PSW mode and the Gaussian field profile.

\begin{figure}[h!]
	\centering
	\includegraphics[width=0.6\linewidth]{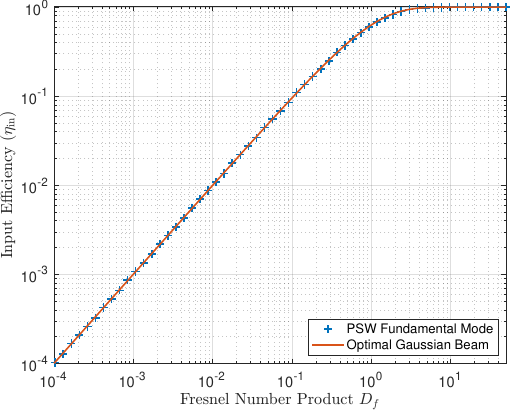}
	\caption{Comparison of the input efficiency $\eta_\mathrm{in}$ for the optimal (numerically optimized $ \xi_0 $ Gaussian input field's transmissivity (blue cross marks) with the fundamental PSW mode $ \varphi_{0,0} $ (red-solid line) w.r.t. $ x=\sqrt{D_f} $. }
	\label{fig:tot_eff}
\end{figure}

{ In addition to the transmission efficiency of the input fields, practical implementations of FSO links also need to account for the overall power transfer efficiency from the source, i.e., fraction of the total input power that is captured by the transmitter's exit pupil. Beam transmissivity as defined in Eq.~\eqref{eq:trans} normalizes the input power, i.e.\ we examine the ratio between the powers at the transmitter and receiver apertures; to study the power transfer efficiency, we  calculate the \textit{input efficiency}, $\eta_\mathrm{in}$, for the transmitter side aperture as,
\begin{align}
	\eta_\mathrm{in}= \frac{\int_{|\boldR|\leq r_T } |u_I(\boldR)|^2 d^2\boldR}{\int_{|\boldR|< \infty } |u_I(\boldR)|^2 d^2\boldR}.
	\label{eq:trans_n}
\end{align}
Correspondingly, the overall efficiency is given as $\eta_\mathrm{in}\times \eta$. We plot this quantity for both the PSW fundamental mode and the optimal Gaussian beam in Fig.~\ref{fig:tot_eff} -- the calculated quantity is noted to be equal for both the beams. }

\section{Conclusion and Discussion}
The prolate spheroidal wavefunction basis provides a rigorous and complete description of free-space optical propagation between finite circular apertures, and its fundamental mode is known to maximize single-mode transmissivity for such channels. However, the practical relevance of this result has remained limited by the mathematical complexity of PSW modes and the absence of a clear connection to experimentally realizable optical fields. This work closes that gap by providing a clear restatement of Slepian's results~\cite{Slepian1964-eh} in modern notation. 

We have shown that, for free-space optical links bounded by hard circular apertures, an aperture-truncated Gaussian beam with an appropriately optimized waist achieves the same transmissivity as the fundamental PSW mode across a broad range of Fresnel number products. This equivalence holds not only in the far field, where Gaussian beams are often expected to perform well, but also deep into the near-field regime, where the PSW mode structure departs significantly from a Gaussian profile. Although the two fields differ in their detailed spatial structure, particularly in the near field, their transmissivity and overall power transfer efficiency are identical.

Our findings indicate that optimal PSW performance is already achievable with simple Gaussian beams—provided their parameters are chosen correctly. In this sense, the PSW formalism serves not as a prescription for mode generation, but as a theoretical benchmark that justifies and quantifies the optimality of familiar optical beams. From a systems perspective, this equivalence alleviates any requirements for specialized beam, shaping or modal engineering beyond standard Gaussian optics for single-spatial-mode coupling in aperture-limited free-space links. This result provides a favorable outlook for long-range classical links and quantum optical channels, where diffraction-limited loss plays a central role in determining achievable capacity, error rates, and entanglement distribution.

The mathematical underpinnings of our results highlight an important distinction between \textit{mode optimality} and \textit{mode uniqueness}. While the fundamental PSW mode is the unique eigenfunction that maximizes spatial energy concentration, the channel transmissivity itself is not uniquely tied to that mode. In regimes where multiple spatial modes approach unit transmissivity, different input fields—including Gaussian beams—can achieve identical performance despite differing modal decompositions. This observation clarifies why Gaussian beams remain effective in high-Fresnel-number systems and provides a principled explanation for their empirical success. 

While field-deployed FSO links are subject to numerous practical limitations—including atmospheric turbulence, pointing and tracking errors, and optical-axis misalignment—the central conclusion of this work remains robust. Optimal single-mode transmissivity in aperture-limited free-space optical channels can be achieved with simple Gaussian beams, thereby firmly linking a foundational theoretical result to practical optical system design.

\cleardoublepage
\appendix
\section{Prolate Spheroidal Wavefunctions}
\label{app:slepian_defs}

\subsection{Definitions and Derivation}
D. Slepian introduced the prolate spheroidal wavefunction (PSW) modal basis in ~\cite{Slepian1961-sl}. The PSW basis functions are solutions $\psi(x_1,x_2)$ to the two-dimensional eigenvalue relation,
\begin{align}
	\int_{y_1^2+y_2^2\leq1 } \exp({i g \left(x_{1} y_{1}+x_{2} y_{2}\right)})\; \psi\left(y_{1}, y_{2}\right) d y_{1} d y_{2}= \alpha \cdot\psi\left(x_{1}, x_{2}\right).
	\label{eq:slepian_rect}
\end{align} 
 Alternative representation of the eigenvalue relation involve a  change to radial coordinates $ (x_1,x_2)\rightarrow (r,\theta); (y_1,y_2)\rightarrow(r',\theta') $ yielding
\begin{align}
	\alpha\, \psi(r, \theta) 
	&=\sum_{-\infty}^{\infty} i^{m} e^{i m \theta} \int_{0}^{1} d r^{\prime} r^{\prime} J_{m}\left({grr}^{\prime}\right) \int_{0}^{2 \pi} d \theta^{\prime} e^{-i m \theta^{\prime}} \psi\left(r^{\prime}, \theta^{\prime}\right).
\end{align}
The admitted solutions are then shown to take the form,
\begin{subequations}
	\begin{align}
		\psi_{0, n}(r, \theta)&=R_{0, n}(r),\quad  n=0,1,2,\dots,  \\
		\psi_{\ell,p}(r, \theta)&=\begin{cases}
			R_{\ell,p}(r) \cos \ell\theta
			\\R_{\ell,p}(r) \sin \ell\theta  
		\end{cases}\;N=0,1,2,\ldots; n=0,1,2,\ldots,
	\end{align}
\end{subequations}
where $N,n$ are the mode indices and $ R_{\ell,p}(r)$ must satisfy the equation,
\begin{align}
	\beta_{\ell,p} R_{\ell,p}(r)=\int_{0}^{1} J_{\ell}\left(c r r^{\prime}\right) R_{\ell,p}\left(r^{\prime}\right) r^{\prime} d r^{\prime},
	\label{eq:PSW_rad}
\end{align}
with $ \alpha_{\ell,p}=2 i^N \pi \beta_{\ell,p} $. By making the substitutions $\gamma_{\ell,p}=\sqrt{c} \beta_{\ell,p}$ and $\varphi_{\ell,p}(r)=\sqrt{r} R_{\ell,p}(r)$
the \textit{symmetric equation} for the PSW basis is defined by,
\begin{align}
	\gamma_{\ell,p} \, \varphi_{\ell,p}(r)=\int_{0}^{1} J_{\ell}\left(c r r'\right) \sqrt{c r r'}\, \varphi_{\ell,p}\left(r^{\prime}\right) d r^{\prime}.
	\label{eq:PSW_symm}
\end{align}
Evaluation of PSW functions however is non-trivial since the functions are typically expressed as convergent sums of Bessel functions, of the general form, 
\begin{align}
	\varphi_{\ell,p}(x)=\frac{1}{\gamma_{\ell,p}} \sum_{j=0}^{\infty} d_{j}^{\ell,p} \frac{J_{\ell+2 j+1}(c x)}{ \binom{\ell+j}{j} \sqrt{c x}},
	\label{eq:PSW_expansionBessel}
\end{align}
where $J_{n} (\cdot) $ is the Bessel function of the $ n $-th order and $ d_{j}^{\ell,p} $ are coefficients of the expansion, which form a three-term recurrence. This recurrence can be obtained by substituting \eqref{eq:PSW_expansionBessel} in \eqref{eq:PSW_symm} and was originally stated in~\cite{Slepian1964-eh} as,
\begin{align}
	c^{2} \gamma_{\ell, j-1}^{(1)} \,d_{j-1}^{\ell,p} 
	+\left[c^{2} \gamma_{\ell, j}^{(0)}+\left(2 j+\ell+\frac{1}{2}\right)\left(2 j+\ell+\frac{3}{2}\right)-\chi\right] d_{j}^{\ell,p} +c^{2} \gamma_{\ell, j+1}^{(-1)}\, d_{j+1}^{\ell,p}=0
	\label{eq:recur1}
\end{align}
where $\chi$ is the eigenvalue for the mode labelled by $\ell,p$ and the coefficients $ \gamma_{\ell,j}^{(\kappa)} ;\, \kappa=-1,0,1$ are given by
\begin{subequations}
	\begin{align}
		\gamma_{\ell, j}^{(1)} &=-\frac{(j+\ell+1)^{2}}{(2 j+\ell+1)(2 j+\ell+2)} \\
		\gamma_{\ell, j}^{(0)} &=\frac{1}{2}\left(1+\frac{\ell^{2}}{(2 j+\ell)(2 j+\ell+2)}\right) \\
		\gamma_{\ell, j}^{(-1)} &=-\frac{j^{2}}{(2 j+\ell)(2 j+\ell+1)}
	\end{align}
\end{subequations}
It is important to note here that the recurrence relation in \eqref{eq:recur1} only specifies the coefficients for the expansion of \eqref{eq:PSW_expansionBessel} and does not contain any dependence on the $ p $ parameter. In the subsequent section, we shall see that the eigenvalue formulation has multiple solutions (generally infinitely many), and the parameter $ p $ determines which eigenvalue would yield the appropriate mode (i.e. $ n=0 $ determines the mode with the highest transmissivity for a given $ \ell $, $ p=1 $ the subsequent mode, and in like manner the rest.)

\subsection{Solving the recurrence relation of~\eqref{eq:recur1}}
\label{subsec:bouwkamp_method}
To proceed with solving this recurrence, let us simplify the notation by relabeling $ d^{\ell,p}_j $ for a given $\ell,p$ as $ a_j $. Additionally, we may expand the $ \gamma $ terms as defined, with the substitution $ \ell+2j+1=k $. Note that when $ j $ increases (decreases) by 1, $ k $ correspondingly increases (decreases) by 2. Hence, the following relations become apparent when we adopt the `compact' notation,
\begin{subequations}
	\begin{align}
		&	\gamma_{\ell,j-1}^{(1)}=\frac{-c^2 (k-j-1)^2}{(k-1)(k-2)}\\
		&	\gamma_{\ell,j+1}^{(-1)}=\frac{-c^2 (j+1)^2}{(k+1)(k+2)}\\
		&	\gamma_{\ell,j}^{(0)}=\frac{c^2}{2}\left(1+\frac{\ell^2}{(k-1)(k+1)}\right)
	\end{align}
\end{subequations}
Adopting the method of Bouwkamp~\cite{Bouwkamp1947-no} which seeks to resolve three term recurrences similar to~\eqref{eq:recur1}, we adopt the following notations
\begin{subequations}
	\begin{align}
		&\alpha_j= \frac{c^2}{2}\left(1+\frac{\ell^2}{(k-1)(k+1)}\right)+\left(k-\frac{1}{2}\right)\left(k+\frac{1}{2}\right) \label{eq:reframa}
		\\ &\beta_j=c^4\times \frac{j^2(k-j-1)^2 }{k(k-1)^2 (k-2)} \label{eq:reframb}
		\\ &\Gamma_j=\frac{c^2 j^2}{k(k-1)}\frac{a_j}{a_{j-1}} \label{eq:reframG}
	\end{align}
\end{subequations}
These choices for defining the terms allows us to rewrite~\eqref{eq:recur1}  as 
\begin{align}
	\beta_j=(\alpha_j-\chi)\Gamma_j-\Gamma_j\Gamma_{j+1}.
\end{align}
This leads us to the continued fraction description of the $ \Gamma_j$'s as
\begin{align}
	\Gamma_j=\frac{\beta_{j}}{\alpha_j-\chi-\Gamma_{j+1}}\Rightarrow\Gamma_j=\frac{\beta_j\;|}{|\alpha_j-\chi}-\frac{\beta_{j+1}\;|}{|\alpha_{j+1}-\chi}-\frac{\beta_{j+2}\;|}{|\alpha_{j+2}-\chi}-\cdots.
	\label{eq:trns1}
\end{align}
Note that our continued fraction notation should be interpreted as
\begin{align}
	\frac{A_1|}{|B_1}-\frac{A_2|}{|B_2}-\frac{A_3|}{|B_3}-(\cdots)\equiv\cfrac{A_1}{B_1-\cfrac{A_2}{B_2-\cfrac{A_3}{B_3-(\cdots)}}}
\end{align}
Similarly, upon rearranging~\eqref{eq:trns1} we get the relation,
\begin{align}
	\Gamma_{j+1}=\alpha_j-\chi-\frac{\beta_{j}}{\Gamma_{j}}\Rightarrow\Gamma_{j+1}=\alpha_j-\chi-\frac{\beta_j\;|}{|\alpha_{j-1}-\chi}-\frac{\beta_{j-1}\;|}{|\alpha_{j-2}-\chi}-\frac{\beta_{j-2}\;|}{|\alpha_{j-3}-\chi}-\cdots.
	\label{eq:trns2}
\end{align}
Setting $ j \mapsto j+1 $ in~\eqref{eq:trns1}, we obtain a transcendental equation by equating the transformed equation to~\eqref{eq:trns2} as,
\begin{align}
	\frac{\beta_{j+1}\;|}{|\alpha_{j+1}-\chi}-\frac{\beta_{j+2}\;|}{|\alpha_{j+2}-\chi}-\frac{\beta_{j+3}\;|}{|\alpha_{j+3}-\chi}-\cdots=\alpha_j-\chi-\frac{\beta_j\;|}{|\alpha_{j-1}-\chi}-\frac{\beta_{j-1}\;|}{|\alpha_{j-2}-\chi}-\frac{\beta_{j-2}\;|}{|\alpha_{j-3}-\chi}-\cdots.
	\label{eq:trnsSol}
\end{align}
Naturally, we have to terminate the LHS of~\eqref{eq:trnsSol} at some point; we do this by setting $ \beta_{j+k} =0$ for some $ k>0 $. This is the $ k$-th-order approximate transcendental equation. Numerical solvers readily solve the $ k $ eigenvalues $ \chi $ quite readily for $ k\sim10 $; in the present study we use Mathematica. Also, it is important to note that the eigenvalues are ordered to correspond to the $ \varphi_{\ell,p}(x) $ PSW modes, i.e., the smallest eigenvalue corresponds to $ \varphi_{\ell,0} $ and so on. 

The term $ \gamma_{\ell,p} $ in~\eqref{eq:PSW_expansionBessel} can be obtained from the following relation,
\begin{align}
	\gamma_{\ell,p}=\frac{c^{\ell+\frac{1}{2}} d_{0}^{\ell,p}}{2^{\ell+1} \Gamma(\ell+2) \sum_{j=0} d_{j}^{\ell,p}}.
\end{align}

\section{Trace Theorem for Free-Space Optical Communications}
\label{app:trace_class}


In our treatment of free-space optical communications, we have considered monochromatic scalar paraxial propagation over a free-space distance $L$ at wavelength $\lambda$.
Let $\psi_t(\rt)\in L^2(\mathbb{R}^2)$ denote the complex field envelope in the transmitter plane and $\psi_r(\rr)$ the field envelope in the receiver plane. The transmitter and receiver pupils are described by aperture (or apodization) functions $a_t(\rt)$ and $a_r(\rr)$, which may be hard (circular or rectangular) stops, soft-Gaussians, or general bounded masks. The physical field that is launched is $a_t(\rt)\psi_t(\rt)$ and the aperture-truncated received field that is sent to for further detection and processing, is $a_r(\rr)\psi_r(\rr)$. The Fresnel input--output relation is written in integral form as
\begin{equation}
\psi_r(\rr)
= \int_{\mathbb{R}^2}\! d^2\rt\;
K_L(\rr,\rt)\, a_t(\rt)\,\psi_t(\rt),
\label{eq:io_integral}
\end{equation}
followed by receiver truncation
\begin{equation}
\phi_r(\rr) = a_r(\rr)\,\psi_r(\rr).
\label{eq:rx_trunc}
\end{equation}
For paraxial propagation in vacuum, one convenient normalization of the Fresnel kernel is
\begin{equation}
K_L(\rr,\rt)
= \frac{1}{i\lambda L}\exp\!\left[\frac{i\pi}{\lambda L}\abs{\rr-\rt}^2\right],
\label{eq:fresnel_kernel}
\end{equation}
for which
\begin{equation}
\abs{K_L(\rr,\rt)}^2 = \frac{1}{(\lambda L)^2}.
\label{eq:kernel_mag}
\end{equation}
The modulus in Eq.~\eqref{eq:kernel_mag} is the only property used in the main trace theorem we derive; quadratic phases affect eigenfunctions but not the trace. Throughout, we assume that 
\begin{equation}
a_t, a_r \in L^2(\mathbb{R}^2)\cap L^\infty(\mathbb{R}^2),
\qquad
0\le \abs{a_{t,r}(\cdot)}\le 1,
\label{eq:aperture_conditions}
\end{equation}
which includes finite-area hard pupils and Gaussian apodizations.
Define the \emph{effective pupil areas}:
\begin{equation}
A_t \coloneqq \int_{\mathbb{R}^2} d^2\rt\,\abs{a_t(\rt)}^2,
\qquad
A_r \coloneqq \int_{\mathbb{R}^2} d^2\rr\,\abs{a_r(\rr)}^2.
\label{eq:areas}
\end{equation}
For hard pupils, $A_{t,r}$ are geometric areas; for apodized pupils they are the $L^2$-norm areas.

For ease of representation, it is convenient to package the channel into a single linear operator $K:L^2(\mathbb{R}^2)\to L^2(\mathbb{R}^2)$ defined by
\begin{equation}
(K\psi)(\rr) \;=\; a_r(\rr)\int d^2\rt\; K_L(\rr,\rt)\,a_t(\rt)\,\psi(\rt).
\label{eq:K_def}
\end{equation}
The input-output relation is thereby denoted: $\phi_r = K\psi_t$. We now note a few properties
\begin{enumerate}
    \item \textbf{Power transfer operator and modal transmissivities}: We may then define the \emph{power-transfer operator} on the input space,
    \begin{equation}
    \mathcal{T} \;\coloneqq\; K^\dagger K.
    \label{eq:T_def}
    \end{equation}
    The operator $\mathcal{T}$ is positive semidefinite and self-adjoint. Its eigenvalues $\{\eta_n\}\subset[0,1]$ (counted with multiplicity) are the channel's \emph{modal power transmissivities}: for an eigenfunction $u_n$, also termed the {\em normal spatial modes} of the propagation kernel, with $\norm{u_n}=1$,
    \begin{equation}
    \eta_n \;=\; \norm{Ku_n}^2 \;=\; {u_n}{\mathcal{T}}{u_n}.
    \end{equation}
    Equivalently, the {\em singular} values of $K$ are $\{\sqrt{\eta_n}\}$.

        \item \textbf{Trace-class property}: Under the aperture conditions \eqref{eq:aperture_conditions}, $K$ is a Hilbert--Schmidt operator and $\mathcal{T}=K^\dagger K$ is trace class. Indeed, the Hilbert--Schmidt norm:
    \begin{align}
    \norm{K}_{\mathrm{HS}}^2
    &= \iint d^2\rr\,d^2\rt\;\abs{a_r(\rr)}^2 \abs{K_L(\rr,\rt)}^2 \abs{a_t(\rt)}^2
    \nonumber\\
    &= \frac{1}{(\lambda L)^2}\left(\int d^2\rr\,\abs{a_r(\rr)}^2\right)\left(\int d^2\rt\,\abs{a_t(\rt)}^2\right)
    \nonumber\\
    &= \frac{A_r A_t}{(\lambda L)^2}<\infty,
    \label{eq:HS_norm}
    \end{align}
    where we used Eq.~\eqref{eq:kernel_mag}. 

    \item \textbf{Integral trace formula and cyclic trace}: Let $\mathcal{T}(\rt,\rt')$ denote the integral kernel of $\mathcal{T}$, with slight abuse of notation. Then, for trace-class $\mathcal{T}$,
    \begin{equation}
    \Tr(\mathcal{T}) = \int d^2\rt\; \mathcal{T}(\rt,\rt).
    \end{equation}
    A direct computation from \eqref{eq:K_def}--\eqref{eq:T_def} yields the standard kernel identity in its integral form:
    \begin{equation}
    \Tr(\mathcal{T})
    = \iint d^2\rr\,d^2\rt\;
    \abs{a_r(\rr)}^2\,\abs{a_t(\rt)}^2\,\abs{K_L(\rr,\rt)}^2,
    \label{eq:trace_kernel_form}
    \end{equation}
    which coincides with $\norm{K}_{\mathrm{HS}}^2$. Because $K$ is Hilbert--Schmidt, both $K^\dagger K$ and $K K^\dagger$ are trace class and the cyclicity of trace applies:
    \begin{equation}
    \Tr(K^\dagger K)=\Tr(KK^\dagger).
    \label{eq:cyclic_trace}
    \end{equation}

\end{enumerate}
Under the aperture conditions \eqref{eq:aperture_conditions}, $K$ is a Hilbert--Schmidt operator and $\mathcal{T}=K^\dagger K$ is trace class (see Appendix~\ref{app:HS_compact_spectrum}). Indeed, the Hilbert--Schmidt norm:
\begin{align}
\norm{K}_{\mathrm{HS}}^2
&= \iint d^2\rr\,d^2\rt\;\abs{a_r(\rr)}^2 \abs{K_L(\rr,\rt)}^2 \abs{a_t(\rt)}^2
\nonumber\\
&= \frac{1}{(\lambda L)^2}\left(\int d^2\rr\,\abs{a_r(\rr)}^2\right)\left(\int d^2\rt\,\abs{a_t(\rt)}^2\right)
\nonumber\\
&= \frac{A_r A_t}{(\lambda L)^2}<\infty,
\label{eq:HS_norm}
\end{align}
where we used Eq.~\eqref{eq:kernel_mag}. Since $K$ is Hilbert--Schmidt, $K^\dagger K$ is trace class and
\begin{equation}
\Tr(\mathcal{T})=\Tr(K^\dagger K)=\norm{K}_{\mathrm{HS}}^2.
\label{eq:trace_equals_HS}
\end{equation}
Having identified these properties of the channel propagation operator $K$, we now present the key property for the power transfer operator,

\textbf{Theorem (aperture-agnostic trace theorem).}
\emph{Let $K$ be the free-space channel operator \eqref{eq:K_def} with Fresnel kernel \eqref{eq:fresnel_kernel}, and let $\mathcal{T}=K^\dagger K$ be the corresponding power-transfer operator. Under \eqref{eq:aperture_conditions}, $\mathcal{T}$ is trace class with eigenvalues $\{\eta_n\}$ satisfying}
\begin{equation}
\sum_{n}\eta_n = \Tr(\mathcal{T}) = \frac{A_t A_r}{(\lambda L)^2}.
\label{eq:main_result}
\end{equation}

\emph{Proof.}
Since $\mathcal{T}$ is trace class, $\sum_n \eta_n = \Tr(\mathcal{T})$. Using \eqref{eq:trace_kernel_form} and the Fresnel modulus \eqref{eq:kernel_mag} gives
\begin{align}
\Tr(\mathcal{T})
&= \iint d^2\rr\,d^2\rt\;
\abs{a_r(\rr)}^2\,\abs{a_t(\rt)}^2\,\frac{1}{(\lambda L)^2}
\nonumber\\
&=\frac{1}{(\lambda L)^2}\left(\int d^2\rt\,\abs{a_t(\rt)}^2\right)\left(\int d^2\rr\,\abs{a_r(\rr)}^2\right)
\nonumber\\
&=\frac{A_t A_r}{(\lambda L)^2}.
\end{align}
\hfill$\square$


Equation~\eqref{eq:main_result} states that the total ``mass'' of the transmissivity spectrum is a \emph{purely geometric invariant} fixed by the link distance and the $L^2$-areas of the pupils. Defining the dimensionless Fresnel-number product
\begin{equation}
D_f \;\coloneqq\; \frac{A_t A_r}{(\lambda L)^2},
\label{eq:D_def}
\end{equation}
the theorem becomes
\begin{equation}
\sum_n \eta_n = D_f.
\label{eq:sum_eta_equals_D}
\end{equation}
In other words, $D$ is the phase-space overlap volume admitted by the two pupils, measured in units of $(2\pi)^2$, and thus acts as an effective count of spatial degrees of freedom.

We consider the special cases of Gaussian pupils to demonstrate this explicitly. For this Gaussian--Fresnel channel, the power-transfer eigenvalues can be indexed by a nonnegative integer order $q=1,2,\dots$ with a linear degeneracy $q$ (arising from the number of LG or HG modes of a given total order). The eigenvalues take the form :
\begin{equation}
\eta_{q,m} = \eta_0^{\,q}, \qquad m=1,2,\dots,q,
\label{eq:geom_spec}
\end{equation}
where the base transmissivity $\eta_0\in(0,1)$ is determined by $D$ via the expression:
\begin{equation}
\eta_0 = \frac{1+2D-\sqrt{1+4D}}{2D}.
\label{eq:eta0_D}
\end{equation}
Equivalently, $\eta_0 = (\sqrt{1+4D}-1)/(\sqrt{1+4D}+1)$. Summing \eqref{eq:geom_spec} over all modes gives
\begin{equation}
\sum_n \eta_n = \sum_{q=1}^\infty \sum_{m=1}^{q} \eta_0^{\,q}
= \sum_{q=1}^\infty q\,\eta_0^{\,q}.
\label{eq:deg_sum}
\end{equation}
Using the identity (for $|x|<1$)
\begin{equation}
\sum_{q=1}^\infty q x^q = \frac{x}{(1-x)^2},
\end{equation}
we obtain
\begin{equation}
\sum_{q=1}^\infty q\,\eta_0^q = \frac{\eta_0}{(1-\eta_0)^2}.
\label{eq:sum_closed_form}
\end{equation}
Substituting \eqref{eq:eta0_D} and simplifying yields
\begin{equation}
\sum_{q=1}^\infty q\,\eta_0^{\,q} = D,
\end{equation}
which matches the statement of the aperture-agnostic trace theorem \eqref{eq:sum_eta_equals_D} stated earlier, as expected.

\section{Optimal Gaussian Beam Evolution}
\label{sec: opti_gauss}
\begin{figure}[h!]
	\centering
	\includegraphics[width=0.7\linewidth]{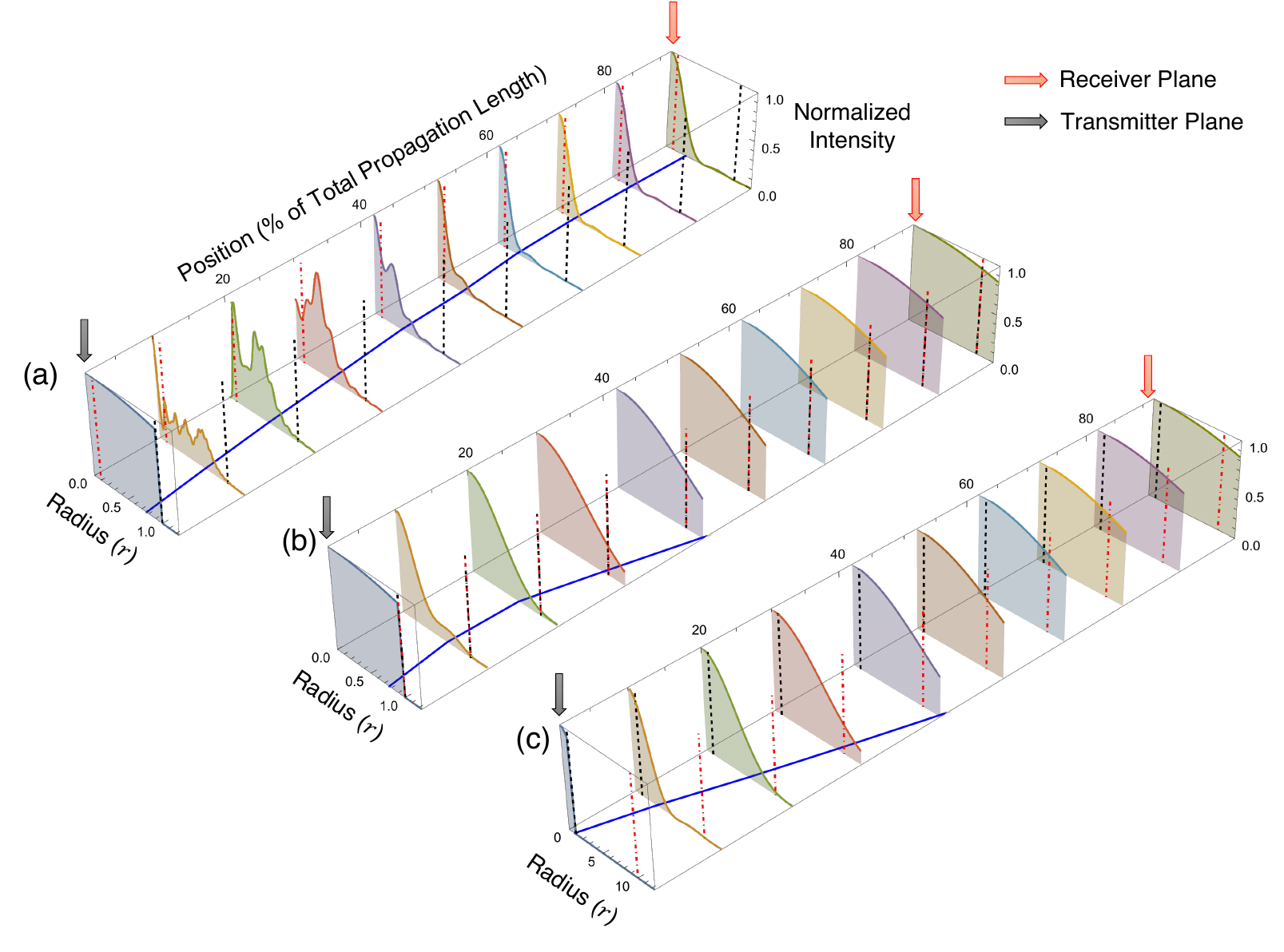}
	\caption{Optimal beam evolution at $ D_f=0.1 $ for $ r_T:r_R $ ratios (a) $1:0.1$ (a) $1:1$ (c) $0.1:1$.  Lines (perpendicular to the $ x-y $ plane) to mark the boundary of the transmitter (black, dashed) and receiver (red, dot-dashed) apertures;  blue line on the bottom $ x-y $ plane to marks the effective `beam width' ( radius at which it captures 60\% of the power in the transverse beam profile) as the beam evolves.}
	\label{fig:df1}
\end{figure}

In this section, we examine the beam width evolution for the optimal Gaussian beam as discussed in the previous section. We will look at the far-field ($ D_f=0.1 $), transition ($ D_f=1 $) and near-field ($ D_f=10 $) regimes for beam propagation, with a different propagation geometries by varying the transmitter-receiver radius pair combinations as \{$ 1\!:\!0.1,\, 1\!:\!1,\, 0.1\!:\!1 $\}.  It is important to note here that limited numerical precision yields beam intensity profiles with large oscillations (see for e.g. Fig.~\ref{fig:df1}(a)).

\begin{figure}[h!]
	\centering
	\includegraphics[width=0.7\linewidth]{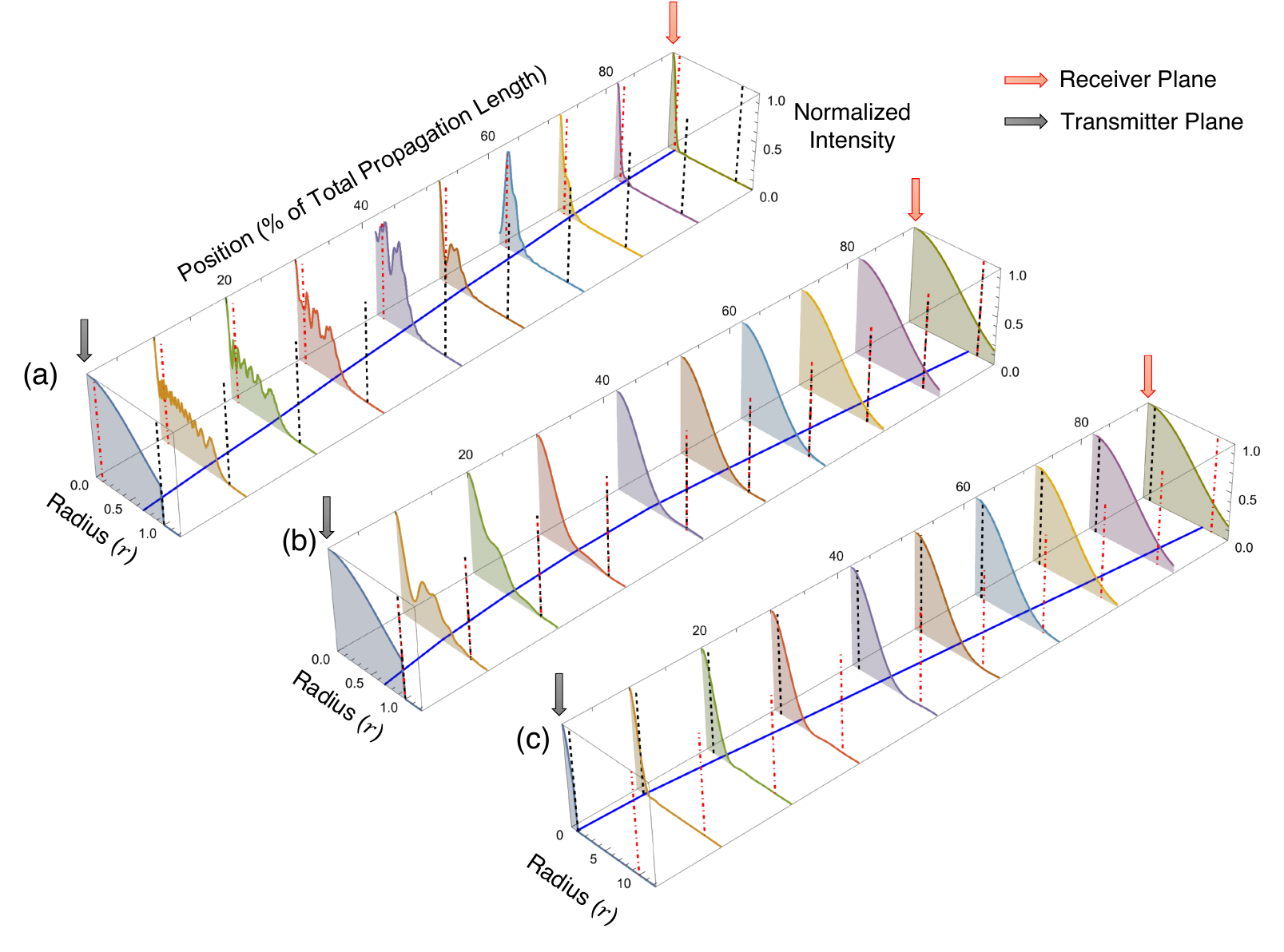}
	\caption{Optimal beam evolution at $ D_f=1 $ for $ r_T:r_R $ ratios (a) $1:0.1$ (a) $1:1$ (c) $0.1:1$. Lines (perpendicular to the $ x-y $ plane) to mark the boundary of the transmitter (black, dashed) and receiver (red, dot-dashed) apertures;  blue line on the bottom $ x-y $ plane to marks the effective `beam width' ( radius at which it captures 60\% of the power in the transverse beam profile) as the beam evolves.}
	\label{fig:df2}
\end{figure}

\begin{figure}[h!]
	\centering
	\includegraphics[width=0.7\linewidth]{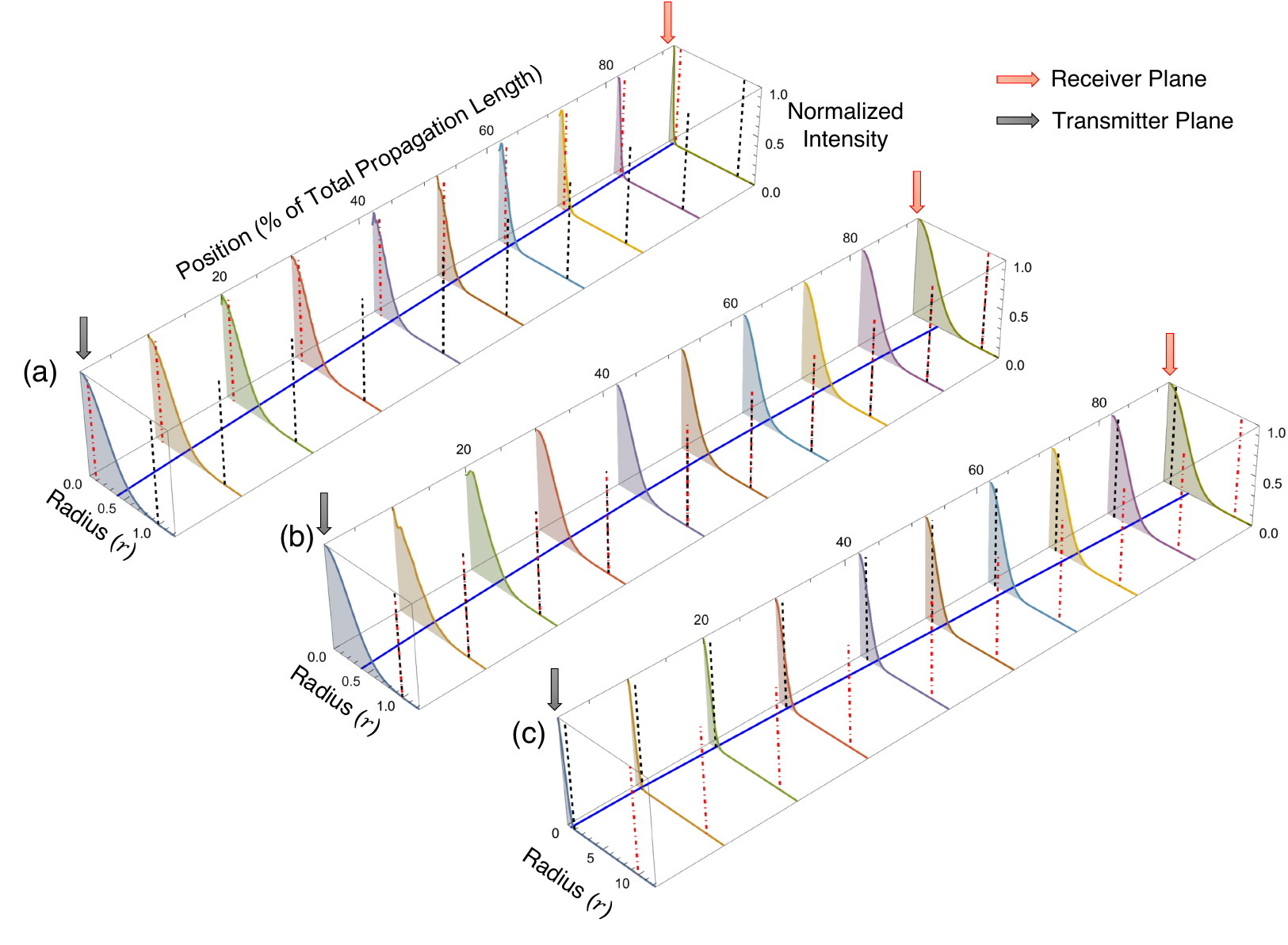}
	\caption{Optimal beam evolution at $ D_f=10 $ for $ r_T:r_R $ ratios (a) $1:0.1$ (a) $1:1$ (c) $0.1:1$. Lines (perpendicular to the $ x-y $ plane) to mark the boundary of the transmitter (black, dashed) and receiver (red, dot-dashed) apertures;  blue line on the bottom $ x-y $ plane to marks the effective `beam width' ( radius at which it captures 60\% of the power in the transverse beam profile) as the beam evolves.}
	\label{fig:df3}
\end{figure}

We depict the beam evolution in  Figs.~\ref{fig:df1}-\ref{fig:df3}. In each figure, we plot the normalized beam intensity ($z$-axis) vs. the radius ($ x $-axis) at various lengths along its propagation length ($y$-axis; marked as \% of the total propagation length). We include additional markers for ease of visualization: for each beam profile, we add lines (perpendicular to the $x-y$ plane) to mark the boundaries of the transmitter (black, dashed) and receiver (red, dot-dashed) apertures. Additionally, we draw a blue line on the bottom $ x-y $ plane to mark the effective `beam width' as the beam evolves. We define the width as the beam radius at which the transverse beam profile captures 60\% of the power.

%
%
%

The main takeaways from the plots are-
\begin{enumerate}
	\item Irrespective of the $ D_f $, the beam evolution shows a loose semblance of a `focus' i.e. we can distinctly see the beam width (blue line) roughly tracking the ratio of the aperture radii.
	\item In the far-field (i.e. $ D_f =0.1$), the beam expands considerably as we track its propagation. In general, the optimal $ \xi_0 $ parameter ensures that the transmitter plane illumination is uniform (`plane-wave' like), resulting in a bright central lobe (similar to the Airy pattern) that fills the receiver aperture to maximize transmissivity.
	\item For near-field geometries (i.e. $D_f=1$), the beam expansion effect isn't distinct. Hence the optimal transmitter plane illumination is a Gaussian distribution that fills the transmitter aperture, which ensures that the receiver aperture distribution is approximately the same.
	\item For geometries in the transition between the near and far-field regime (i.e. $ D_f=1 $), the effects from regimes are visible, i.e. the transmitter aperture illumination is a Gaussian distribution that overfills the transmitter aperture, which leads to the receiver plane intensity profile to also resemble an overfilled Gaussian.
\end{enumerate}

%

\bibliography{PSW_paper_main}

\end{document}